%% file: paper.tex
\documentclass[journal]{vgtc}

\usepackage{mathptmx}
\usepackage{graphicx}
\usepackage{times}
\usepackage{listings}
\usepackage{amsmath}
\usepackage{subfigure}
\usepackage[draft]{pgf}
\usepackage{xspace}
\usepackage{amssymb}
\usepackage{layouts}
\usepackage{epstopdf}
\usepackage{enumitem}
\usepackage{boxedminipage}
\usepackage[innercaption]{sidecap}

\usepackage[bookmarks,backref=true,linkcolor=black]{hyperref} 
\hypersetup{
  pdfauthor = {},
  pdftitle = {},
  pdfsubject = {},
  pdfkeywords = {},
  colorlinks=true,
  linkcolor= black,
  citecolor= black,
  pageanchor=true,
  urlcolor = black,
  plainpages = false,
  linktocpage
}

\newcommand{\Rspace}        {{\mathbb R}}
\newcommand{\Dspace}        {{\mathbb D}}
\newcommand{\Fspace}        {{\mathbb F}}

\newcommand{\hidecomment}[1]{}

\newcommand{\reffig}[1]{{Fig.~\ref{#1}}}

\newcommand{\ie}{\emph{i.e.}\xspace}
\newcommand{\eg}{\emph{e.g.}\xspace}
\newcommand{\etc}{\emph{etc}\xspace}

\newcommand{\myparagraph}[1]{\vspace{0.05cm}\noindent \textbf{#1}}

\onlineid{217}

\vgtccategory{Research}

\title{Urban Pulse: Capturing the Rhythm of Cities}

\author{Fabio Miranda, Harish Doraiswamy, \textit{Member, IEEE}, Marcos Lage, Kai Zhao, Bruno Gon\c calves, \\ Luc Wilson, Mondrian Hsieh, and Cl\'{a}udio T. Silva, \textit{Fellow, IEEE}}
\authorfooter{
\item F. Miranda, H. Doraiswamy, K. Zhao, B. Gon\c calves, and C. Silva are with New York University.
  E-mail: \{fmiranda, harishd, kai.zhao, bgoncalves, csilva\}@nyu.edu
\item M. Lage is with Universidade Federal Fluminense. 
  E-mail: mlage@ic.uff.br
\item L. Wilson and M. Hsieh are with Kohn Pedersen Fox Associates PC. E-mail: \{lwilson, mhsieh\}@kpf.com.
}

\shortauthortitle{Submission 217}

\graphicspath{{./figs/}{./}}

\abstract{Cities are inherently dynamic. Interesting patterns of behavior 
typically manifest at several key areas of a city over multiple temporal resolutions. Studying these patterns can greatly help a variety of experts ranging from city planners and architects to human behavioral experts.
Recent technological innovations have enabled the collection of enormous amounts of data that can help in these studies. However, techniques 
using these data sets typically focus on understanding the data in the context of the city, thus failing to capture the dynamic aspects of the city. The goal of this work is to instead understand the city in the context of multiple urban data sets. 
To do so, we define the concept of an ``urban pulse" which  captures the spatio-temporal activity in a city across multiple temporal resolutions. The prominent pulses in a city are obtained using the topology of the data sets, and are characterized as a set of beats. The beats are then used to analyze and compare different pulses.
We also design a visual exploration framework that allows users to explore the pulses within and across multiple cities under different conditions. 
Finally, we present three case studies carried out by experts from two different domains that demonstrate the utility of our framework.}

\keywords{Topology-based techniques, urban data, visual exploration.\vspace{-0.2cm}}

\CCScatlist{ 
  \CCScat{K.6.1}{Management of Computing and Information Systems}%
         {Project and People Management}{Life Cycle};
         \CCScat{K.7.m}{The Computing Profession}{Miscellaneous}{Ethics}
}

\teaser{
\centering
\includegraphics[width=14cm]{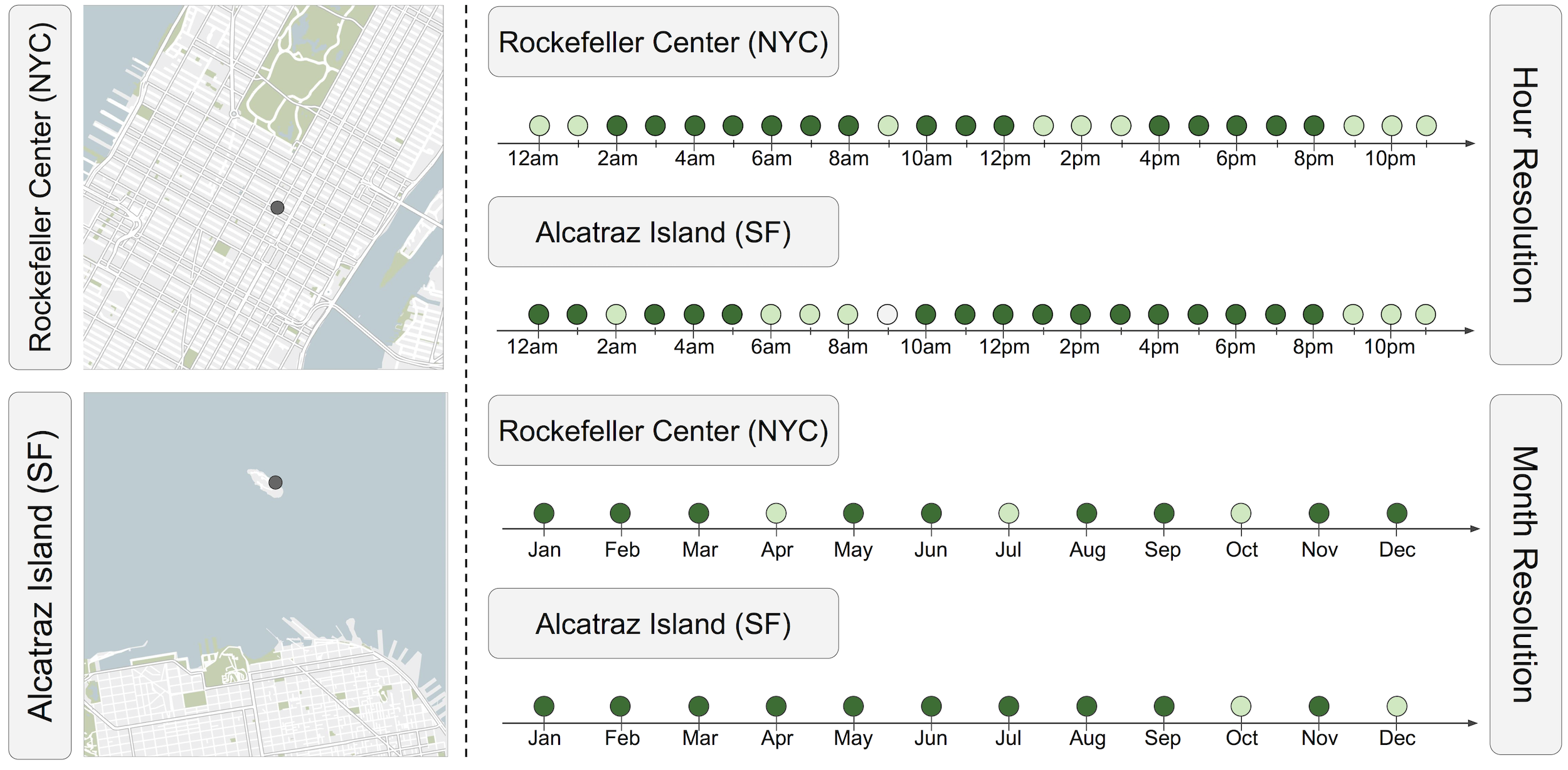}
\vspace{-0.15in}
\caption{Comparing two popular tourist locations -- Rockefeller Center in New York City~(NYC) and Alcatraz Island in San Francisco~(SF), using the \textit{pulse} of these locations. The pulse, defined by a set of \textit{beats} over multiple resolutions, captures the level of activity at a given location. In this example, the beats for the hourly and monthly resolutions are shown, based on Flickr activity. They are computed based on the topology of the time-varying scalar function that models the spatio-temporal distribution of the activity corresponding to a city. 
Dark green represents a significantly high activity at the location, while light green represents a relatively high activity compared to its neighboring locations.
The similarity between the pulses of the two locations over the different resolution indicates that the level of activity is similar across time steps and resolutions even though one is located on the mainland, while the other is an island.}
\label{fig:teaser} 
\vspace{-0.1in}
}

\begin{document}


\firstsection{Introduction}
\label{sec:intro}

\maketitle
\input{intro}

\input{related}

\input{background}
\input{pulse}
\input{impl}
\input{interface}

\input{application}

\input{conclusion}

\acknowledgments{This work was supported in part by a Google Faculty Award, IBM Faculty Award, Moore-Sloan Data Science Environment at NYU, NYU Tandon School of Engineering, NYU Center for Urban Science and Progress, Kohn Pedersen Fox Associates, AT\&T, NSF awards CNS-1229185, CCF-1533564 and CNS-1544753, CNPq, and FAPERJ.}


\bibliographystyle{abbrv}
\bibliography{paper,topology}
\end{document}

%% file: intro.tex
Cities have been the loci of economic activity and human development through the ages, and will continue to be the source of many of the innovations and solutions to the challenges faced by society.
In less than a century, the world has moved from less than $20\%$ of the population residing in urban areas to over $50\%$. A number that is expected to rise to $70\%$ by $2050$, and up to $90\%$ in North America alone~\cite{unurban}.
%
%
This rapid increase in urbanization gives rise to several challenges on how to efficiently transport, house, educate, employ and even entertain an ever increasing number of citizens on a daily basis. A better understanding of the structure of the population, and on how it is distributed and evolving can greatly help in this endeavour.

Technological innovations have enabled the automatic collection of a diverse set of data about our daily lives both as individuals and as a society. 
As a consequence, cities are not only collecting, but also making unique data sets available through open data portals and live feeds~\cite{opendata@book2013,opengov,OpenGovDataUK2012}. An example that has quickly been followed by several private firms~\cite{twitter,flickr} that understand the power of open data and of the potential that can be unlocked by combining diverse sources of data. 
%
This opens up new opportunities for city governments and social scientists to engage in data-driven science to better understand cities, and improve the lives of their residents.
Not surprisingly, there have been several techniques proposed in recent times with the above goal in mind (\eg see \cite{Andrienko2013c,Andrienko2013a,doraiswamy@tvcg2014,ferreira@vast2015,ferreira2013visual,poco@eurovis2015}).
However, all of these techniques focus on identifying features or events in the data, or exploring the mobility of predefined entities. That is, the focus in these works was on the data in the context of a city.  

Our goal in this work is \textit{to understand the city in the context of the different data sets}, inspired by the theory propagated by Park and Burgess in their seminal work \textit{The City}~\cite{park1925city}. They put forth the idea that a city ``is involved in the vital processes of the people who compose it, and is a product of nature and particularly of human nature". 
%
They propose an analogy between the process occurring within a city and the heart beat or \textit{pulse of a human body}. Here, the processes are defined by empirical measures of human activity that vary and grow over time: telephone calls, letters sent by mail, horse-car lines, etc. Unlike Park's time period (the year 1925), we now have several data sets from cities which can help in the identification and analysis of the \emph{pulse of cities}.
In particular, data sets that capture the presence and movement of people (or entities related to people) can be used as proxy to inform the activities in a city. For example, tourist activity is mirrored through Flickr photos, and the general movement of people in a city can be captured by transportation data such as taxis, buses, and subways.

\hidecomment{
It is of vital importance for the various stakeholders of a city to understand its patterns and processes.
For instance, an urban planner designing a public space frequently looks for precedent designs that can be used to guide the design of the planned space. The similarity is typically based either on prior experience, or on physical characteristics such as size of the space, or its function (\eg whether it is a park in a residential area, a transit center, \etc.). More important than the intentions of the original planner are ways in which citizens use it. Using empirical data on people's behavior obtained through multiple urban data sets enables us not only to help define new types of public space but will also lead to better functioning and designed public spaces.}

\myparagraph{Pulse of a city.}
Our goal is to capture both the spatial and temporal variation of the ``activity" in a city from such data sets, which we aim to represent as a collection of \textit{pulses}. Similar to a pulse in the human body, the pulse of a city can be observed at several locations in the city -- strong in some locations, weak in some, and cannot be felt in others. The characteristic of the pulse at a given location, whether it is strong or not and how it varies in time, defines the nature of that location. 
It is of vital importance for the various stakeholders of a city to understand such patterns and processes. For instance, an urban planner designing a public space frequently looks at precedent examples that can be used to guide the design of the planned space. 
Using empirical data on people's behavior obtained through multiple urban data sets enables us not only to help define new types of public spaces but will also lead to better functioning and designed public spaces.

However, the problem of identifying the set of pulses corresponding to a city poses several challenges. 
First and foremost, we need to suitably define the notion of a pulse in a city so that it captures both the spatial and temporal characteristics of the city.
Second, since no part of a city is typically inactive for extended periods of time, we must be able to differentiate between locations having strong and weak pulses. 
Third, the set of pulses can have varying characteristics depending on the time steps and temporal resolution. For example, consider New York City~(NYC). Williamsburg, a neighborhood full of restaurants and bars, will have a lot of activity during the later part of the day, while Central Park might be more active during the day than night. At a different temporal resolution, parks might be more active during summer than in winter, while Williamsburg  will be equally active during all seasons. 
It is therefore important that the pulses in a city are defined such that they capture this variation over multiple temporal resolutions.
%
%
Furthermore, users must also be able to compare not just pulses but also cities based on their pulses. Such comparison, for example, can be used by urban planners to identify appropriate precedents from well developed cities when designing a new public space.

\myparagraph{Contributions.}
In this paper, we define the concept of an \emph{urban pulse} which summarizes the spatio-temporal activity in a city across multiple temporal resolutions. We use a data-driven approach where the available spatio-temporal urban data sets determine the pulses of a city.
Our approach combines techniques from computational topology with visual analytics to efficiently identify, explore and analyze pulses across cities.
This is accomplished by first modeling the urban data as a collection of time-varying scalar functions over different temporal resolutions, where the scalar function represents the distribution of the corresponding activity over the city.
The topology of this collection is then used to identify the locations of prominent pulses in a city.

We then characterize the pulse at a given location as a set of \textit{beats} which captures the level of activity at that location over different time steps and resolutions. The beats of the pulses are then used to analyze and compare pulses.
The urban pulse framework includes a visual interface that can be used to explore pulses within and across multiple cities, and to also compare multiple settings. For example, users can compute and compare pulses across different seasons (summer vs. winter), or based on time of day (day vs. night), \etc.

Working in collaboration with domain experts, we demonstrate the utility of the urban pulse framework through multiple case studies:
\vspace{-0.2cm}
\begin{enumerate}[leftmargin=*,noitemsep]
    \item Architects and urban designers from a leading firm used the urban pulse framework in two case studies to help identify precedents for urban design and to better understand neighborhoods. 
    \item An expert specialized in the data driven study of human behavioral patterns used our framework to better understand how the different cultural communities that constitute NYC share its spaces.
\end{enumerate}
\vspace{-0.2cm}
Feedback from the experts suggests that our technique leads to a new understanding and characterization of an urban environment.

%% file: related.tex
\section{Related Work}
\label{sec:related}

Visual analysis approaches targeting urban data have received a lot of attention recently due to the rapid increase in the availability of data from cities.
In this section, we restrict ourselves to approaches that handle spatio-temporal data sets.

There have been a lot of visual analysis tools that target specific data sets and allow users to freely explore the data at various levels of aggregation~\cite{Andrienko2013c,sun2013survey}. 
Similarly, techniques to visually explore object movement include using density-map based visualizations~\cite{SW11,WV09}, and information visualization techniques~\cite{ferreira2013visual,wang2013visual}.
Several systems have also been proposed to visualize human mobility patterns based on social media data~\cite{chen2016interactive,von2016mobilitygraphs}, telephone data~\cite{sbodio2014allaboard,wu2016telcovis}, and public transport data~\cite{palomo2016visually,yu2015iviztrans}.
Andrienko~et~al.~\cite{Andrienko2013} surveyed other techniques common in the visual analytics of movement data.
More recently, Poco~et~al.~\cite{poco@eurovis2015} modeled traffic movement as a vector function, and adapted vector field techniques to visualize flow of traffic in a city.
Alternate techniques to analyze urban data include those that identify features or events in order to summarize the properties of the data. 
Maciejewski~et~al.~\cite{maciejewski2010visual} modeled hotspots in spatial distributions to detect anomalous hotspots.
Andrienko~et~al.~\cite{Andrienko2013a} proposed visual analytics procedures to determine places of interest based on the occurrence of events.
While these techniques are extremely useful for studying the data set or property of interest, getting a sense of their effect on a city requires manually keeping track of the features through an exhaustive exploration of the data, and becomes impractical due to the data sizes (which often span several years).
%
%
Our technique, on the other hand, provides a concise summary of the spatio-temporal variations, allowing users to quickly gauge change in activity across both space and time, and to easily compare locations within as well as across cities in a data set agnostic manner.

\begin{figure*}
    \centering
    \includegraphics[width=0.97\linewidth]{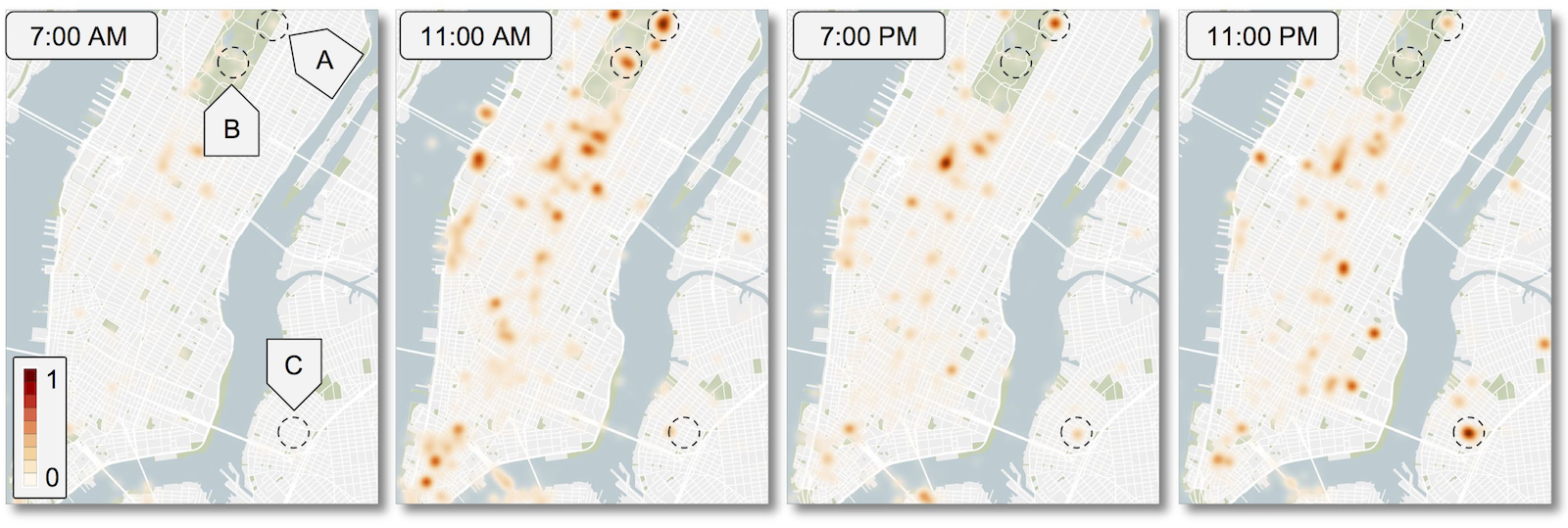}
    \vspace{-0.2in}
    \caption{Density functions computed from the Flickr data set at different time steps. The functions were computed for the hourly time resolution. Observe that there is always activity, indicated by a locally high function value, in the region corresponding to the Metropolitan Museum~(Location A). Central Park~(Location~B), on the other hand, has a relatively high level of activity only during the day.  A neighborhood in the Williamsburg region exhibits~(Location~C) the opposite behavior, being active during the night.}
    \label{fig:scalar-fn}
    \vspace{-0.2in}
\end{figure*}
There are several approaches that use urban data to study different properties of a city. Zheng et al.~\cite{Uair} used taxi information to monitor the air pollution in a city. Zhao~et~al.~\cite{ICDMW15} used recurring spatio-temporal patterns to classify cities into functional regions. 
Zhang~et~al.~\cite{zhang2014visual} proposed a visual analytics system to study patterns of complaints in a city. Claudio~et~al.~\cite{claudio2014metro} combined tourist reviews with transit data for tour planning.
Social media data have been used to develop tools to visualize trends that help get a sense of human activity~\cite{mckenzie2014poi,xia2014citybeat} in cities, to study transport systems~\cite{itoh2014visual}, and to understand land usage patterns~\cite{quercia2014mining}. They have also been used to compare cities based on human movement patterns~\cite{humandiffusion,taleofcities}.
These approaches are typically fine-tuned to work with the targeted data set making them difficult to generalize across a broad spectrum of spatio-temporal urban data. This also makes it difficult to use these tools to compare the properties of a city with respect to multiple data sets and resolutions.

Topology-based techniques are naturally suited for the analysis of data involving spatial and geometric domains.
They have been commonly used to visualize and better understand time-varying data sets in scientific domains (\eg see~\cite{Bremer10,DNN13,Pascucci11}).
More recently, Doraiswamy~et~al.~\cite{doraiswamy@tvcg2014} applied these techniques on spatio-temporal urban data sets to identify events across time and used these events to guide users in the data exploration process.
Topology-based techniques are ideal for our purpose due to two main reasons -- they can efficiently identify locations of interesting features over a spatial domain; and their ability to abstract the data into different classes (\eg based on critical points) allows a simple yet insightful representation of the characteristics of the data (see Section~\ref{sec:pulse-beats}).

%% file: background.tex
\section{Background}
\label{sec:bkgd}

A key step in identifying the set of pulses of a city is to locate
the regions corresponding to the significant pulses. Here, the pulses are defined with respect
to an activity of interest associated with the urban data set that is being
considered. Given an activity of interest, one way to identify the set of 
important locations is to look at regions where that activity is significantly 
more prominent compared to its neighboring regions. 
Topology-based techniques naturally capture such behavior and provide efficient algorithms to represent and compute them. In this section, we briefly introduce the required mathematical background, which are based on concepts from computational topology. We refer the reader to the following textbooks~\cite{EH09,morsebook} for a comprehensive discussion on these topics.

\myparagraph{Modeling data as scalar functions over time.}
A \emph{scalar function}, $f:\Dspace \rightarrow \Rspace$, maps points in a spatial domain $\Dspace$ to real values. In this work, we are interested in the spatial region corresponding to a city, which is represented by a planar domain. A function value is defined over each point in this planar domain with the goal to
capture the activity at that location corresponding to a given data set. 

An example of a scalar function, that is used in this work, is the \emph{density function}. Assume that the input data is provided as a set of points (data points) having location and time. The density function at a given location $p$ is defined as the Gaussian weighted sum
\vspace{-0.1in}
\begin{equation}
\small
f(p) = \sum_{x_i \in N(p)} e^{\frac{-d(p,x_i)^2}{\epsilon^2}}    
\vspace{-0.1in}
\end{equation}
of data points $x_i$ that are in its neighborhood $N(p)$.
Here $d()$ is the Euclidean distance between two points, and $\epsilon$ is the extent of the influence region for a given data point. 
The neighborhood $N(p)$ is defined as a circular region centered at $p$ (see Section~\ref{sec:impl}).

Intuitively, the density function captures the 
level of activity over different locations in a city. For example, consider data corresponding to Flickr images~\cite{flickr}. Each data point corresponds to an image and provides the location where the image was taken together with the time when it was taken. A high function value at a given location implies a lot of activity (there are many pictures being taken) implying the popularity of that location. 

In order to efficiently compute the topological features of the scalar function $f$, it is represented as piecewise linear~(PL) function $f:K \rightarrow \Rspace$.
The planar domain $\Dspace$ of the function is represented by a triangular mesh $K$.
The function is defined on the vertices of the mesh and linearly interpolated within each triangle.

To take into account the variation with time, the set of data points
are first grouped together into a discrete set of time steps corresponding to the temporal resolution, and the scalar function is computed for each of these time steps. 
Again, consider the Flickr example. 
To study the activity at different times of a day, the data is first grouped into 24 time steps corresponding to hourly intervals. The density function is then computed for each of these time steps using the corresponding group of data points. 
\reffig{fig:scalar-fn} illustrates the density function computed on the Flickr data at different times of the day with respect to NYC.

\myparagraph{Critical points.}
The \emph{critical points} of a smooth real-valued function are exactly where the gradient becomes zero. Points that are not critical are \emph{regular}.
The critical points of a PL function are classified based on the behavior of the function within a local neighborhood, and are always located at vertices of the mesh~\cite{Ban70,EHNP03}. Here, the local neighborhood of a vertex is defined using the \emph{link} of that vertex. The link of a vertex $v$ is defined as the mesh induced by the vertices adjacent on $v$. The \emph{upper link} of $v$ is the mesh induced by adjacent vertices having function value greater than $v$, while the \emph{lower link} of $v$ is the mesh induced by adjacent vertices having function value lower than $v$.
A simulated perturbation of the function~\cite{Ede01} imposes a total order on the vertices of the mesh, allowing an unambiguous comparison between vertices.

Critical points are characterized by the number of connected components of the lower and upper links of the vertices. A vertex is \emph{regular} if it has exactly one lower link component and one upper link component. 
A vertex is a \emph{maximum} if its upper link is empty and a \emph{minimum} if its lower link is empty. All other critical points are \emph{saddles}.

\myparagraph{Topological persistence of critical points.}
Given a scalar function $f$, a \emph{super-level set} of a real value $a$ is defined as the pre-image of the interval $[a, +\infty)$. That is, it is the set of all points having function value greater than or equal to $a$ and consists of zero or more connected components.
Consider a sweep of the function $f$ in decreasing order of function value. We are interested in the evolution of the topology of super-level sets against decreasing function value. 
Topological changes occur at critical points, whereas topology is preserved across regular points~\cite{morsebook}. 
In particular, a new super-level set component is created when the sweep passes a maximum, and an existing super-level set component is destroyed at a saddle.

A critical point is called a creator if a new component is created, and a destroyer otherwise. A creator $v_c$ can be uniquely paired with a destroyer $v_d$ that  destroys the component created at $v_c$~\cite{ELZ02}.
The persistence value of $v_c$ is defined as $\pi_c = f(v_c) - f(v_d)$, which intuitively determines the lifetime of the feature created at $v_c$, and is thus a measure of the importance of $v_c$. The traditional persistence of the global maximum is equal to $\infty$ since there is no pairing destroyer for that maximum. In this paper, we use the notion of extended persistence~\cite{AEHW06} which pairs the global maximum with the global minimum. 
Topological persistence has been shown to be an effective importance measure, and has been used in many applications (\eg see \cite{BEHP04,DLLRSY10,GNPB06}).
%
%
Given an input mesh of size $n$, the persistence of the set of maxima can be computed efficiently in $O(n \log n)$ time~\cite{AEHW06}. 

Locations of interest with respect to a given scalar function $f$ are identified using the set of high persistent critical points of $f$.

%% file: pulse.tex
\section{Urban Pulse}
\label{sec:pulse}

Our goal is to capture the spatio-temporal variation of the activity in the city. To this end, we define a pulse as follows:

\myparagraph{Definition.}
A pulse is formally defined as a pair $P = (L,B)$, where $L = (x,y)$ denotes the location of the activity, and $B$ representing the beats that summarizes the variation of that activity over different temporal resolutions at the specified location. 

While every location in a city has a corresponding pulse defined with respect to it, we are interested in the set of prominent pulses -- locations where the beats of the pulse are \emph{stronger}.
The rest of this section focuses on identifying these locations (Section~\ref{sec:pulse-loc}), defining and computing the beats (Section~\ref{sec:pulse-beats}), and quantifying the pulses (Section~\ref{sec:pulse-analysis}).
For ease of exposition, we will be using the density function computed using the Flickr data to illustrate our technique. Note that our technique is oblivious to the scalar function used.

\subsection{Identifying Pulse Locations}
\label{sec:pulse-loc}

The level of activity at a given location can significantly vary depending on the time period in which it is considered. For instance, consider the example in \reffig{fig:scalar-fn} which shows the variation of the density function at different times of the day. Certain locations like Metropolitan Museum~(Met) (Location~A) are always bustling with activity, while other locations such as parks (Location~B) or locations of popular bars in the Williamsburg region~(Location~C) are active only during certain times of the day. Moreover, the level of activity at these locations changes when considered at a different temporal resolution. For example, the Williamsburg area is always active during all months of the year.
It is thus important to capture the variation of a given activity at multiple temporal resolutions. To do this, we first define multiple time-varying scalar functions corresponding to the different resolutions. 

\myparagraph{Handling multiple temporal resolutions.}
In this paper, we restrict the functions to be a subset of the following four time resolutions:

\vspace{-0.2cm}
\begin{itemize}[leftmargin=*,noitemsep]
    \item \textbf{All:} A single scalar function is used to represent the entirety of the data.
    \item \textbf{Month of Year:} There is one scalar function computed for data corresponding to each month of the year. For example, when using the Flickr data, the data points are first grouped by the month in which the picture was taken.  Twelve density functions are then computed using the data points corresponding to each month.
    \item \textbf{Day of Week:} There is one scalar function computed corresponding to each day of the week resulting in a total of seven functions.
    \item \textbf{Hour of Day:} The data is grouped based on the time of day, and one function is computed corresponding to each hour resulting in twenty four functions.
\end{itemize}
\vspace{-0.2cm}
Given a time resolution, the scalar functions corresponding to that resolution are normalized based on the maximum function value over all functions in that resolution. This allows for a consistent comparison of pulses not only between resolutions, but also between time steps within the same resolution.

\myparagraph{Prominent pulse locations.}
Let $Res$ denote the possible resolutions for a given data set (see Section~\ref{sec:impl} for details). For the Flickr data used so far, $Res = \{All, Month, Day, Hour\}$. Let $\Fspace_{r}$ denote the collection of scalar functions corresponding to the different time steps of resolution $r \in Res$. Let $\Fspace = \bigcup \Fspace_r$ over all resolutions in $Res$. Let $C^+_i$ denote the set of high persistence maxima of a scalar function $f_i$. 
We define a maximum to be high-persistent if its persistence value is above a given threshold. Assuming the function values are normalized between 0 and 1, we use a threshold of $0.2$ in this paper.
Let $v$ be a vertex in the input mesh corresponding to a location in a city. We say this location is \textit{prominent} if there exists a scalar function $f_i$ in $\Fspace$ such that $v \in C^+_i$, \ie $v$ is a high persistent maximum of the function $f_i$. 
In other words, a prominent location is a high persistent maximum of a function corresponding to \emph{at least} one time step over all resolutions. 

The highlighted locations in \reffig{fig:scalar-fn} represent some of the prominent pulse locations in NYC. Of these, the location corresponding to the Met is a high persistent maximum from 11~am to 11~pm, while the one corresponding to Williamsburg is a high persistent maximum only during the later part of the day.

\begin{SCfigure}[0.6]
    \centering
    \includegraphics[width=0.55\linewidth]{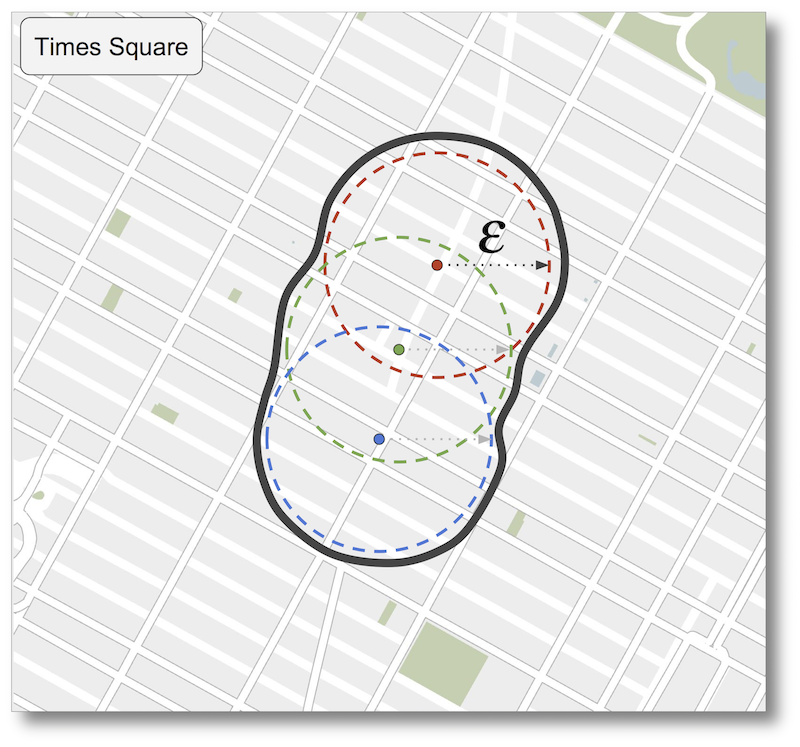}
    \caption{The actual location of the maximum corresponding to the Times Square region, shown in red, green and blue, varies slightly over time.  However, since the variations are within an $\epsilon$-distance, they are combined to represent the region.} \label{fig:pulse-merge}
    \vspace{-0.2in}
\end{SCfigure}

It is possible that the location of a maximum varies slightly along different time steps and / or resolutions.  In order to identify these locations as the same, we cluster the set of prominent locations such that two locations are considered to be the same if they are within an $\epsilon$-distance of each other. 
%
%
For example, the location of the maximum corresponding Times Square has a slight variation since people take pictures along a stretch of the road (red, green and blue dots in \reffig{fig:pulse-merge}). Since all of these locations are within the $\epsilon$ influence radius of another, we combine them to represent that location (black region in \reffig{fig:pulse-merge}).
Here, $\epsilon$ is the user specified influence region of a data point. It is the same as the one used for computing the density function described in Section~\ref{sec:bkgd}. 
\subsection{Capturing Pulse Beats}
\label{sec:pulse-beats}

Consider the locations identified in the previous step. These locations typically have different behavior at different times as seen previously with the Met and Williamsburg in NYC. Some always have high activity (a high persistent maximum), while others are high persistent only during select time periods. Moreover, some locations might not even be a maximum at certain times.
%
%
We use this observation to define three types of \textit{beats} to capture such variations over time.

\begin{figure}
    \centering
    \includegraphics[width=\linewidth]{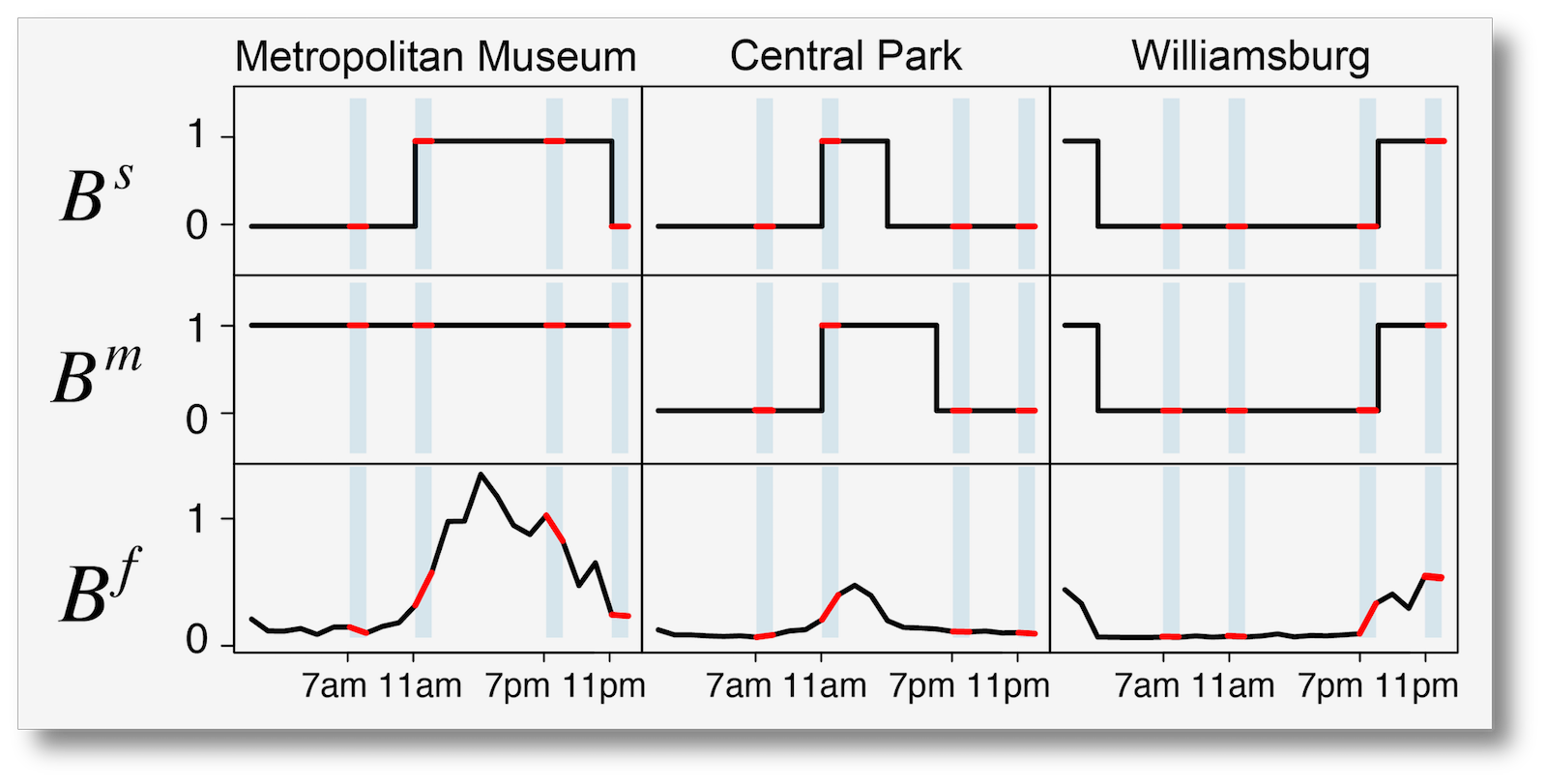}
    \vspace{-0.3in}
    \caption{The three types of beats -- Significant beats ($B_S$), Maxima beats ($B_m)$, and Function beats ($B_f$) for the locations highlighted in \reffig{fig:scalar-fn}. The  time steps corresponding to the ones in  \reffig{fig:scalar-fn} are highlighted.}
    \label{fig:pulse-beats}
    \vspace{-0.2in}
\end{figure}

\vspace{-0.20cm}
\begin{itemize}[leftmargin=*,noitemsep]
    \item \textbf{Significant Beats $B^s$:} This is a 0/1 sequence indicating the absence / presence of a high persistent maximum at that location over the different time steps corresponding to the temporal resolution. This time series gives the user an indication as to when a given location becomes prominent. Capturing the significant beats over a given temporal resolution allows for better utilization of space and infrastructure relative to actual occupation. 
    \item \textbf{Maxima Beats $B^m$:} This is a 0/1 sequence indicating the absence / presence of a maximum at the location over the different time steps. This time series indicates how often a particular location is interesting. For example, a location could be a high persistent maxima only during one time step, but could still be a maxima in several time steps indicating that the location, while being prominent only occasionally, could still be of interest locally. For example, consider the region of Central Park highlighted in \reffig{fig:scalar-fn} (Location~B). 
    While this location has a high level of activity (pictures taken) primarily in the afternoon, one notices that there is still some activity, albeit not very high (it is still a maximum), during day time. 
    Note that this area houses the Rumsey Playfield, a picturesque venue, known for staging numerous free entertainment performances.
    The Met, is a maximum throughout the day, even though it is significant when it is open. This implies that there is still some activity at that location when it is closed. This is because, in addition to being an attraction for its art collection, the Met is also architecturally significant and prominently featured in popular media. 
    \item \textbf{Function Beats $B^f$:} This is a time series showing the variation of the scalar function at that location. If the location encompasses multiple vertices of the mesh, then the maximum function value over all these vertices is used for the time series. This time series allows the analysis of the scalar function corresponding to a location.
\end{itemize}
\vspace{-0.15cm}
Note that all the above beats are defined for each of the possible temporal resolutions, and represent the scalar functions, which aggregate the data, and not the raw data itself.
\reffig{fig:pulse-beats} shows the different beats corresponding to the locations and time steps shown in \reffig{fig:scalar-fn}.

\subsection{Pulse Analysis}
\label{sec:pulse-analysis}

%
Depending on the data set being used there can be a large number of pulses in a city. It is therefore important to rank and compare various pulses to help users in the exploratory analysis. To accomplish this, we first create a feature vector corresponding to each pulse. The feature vectors are then used for ranking and comparison.

\myparagraph{Pulse Rank.}
Consider a pulse $P$. We construct a $d = 3 \times |Res|$ dimensional feature vector $\textbf{p} = (p_1,p_2,\ldots,p_d)^T$ associated with the pulse $P$. Each dimension of the feature vector corresponds to a beat of the pulse. Recall that there are three beats computed for each resolution.
Let a beat $B_x = \{b_1, b_2, \ldots, b_t\}$, where $t$ is the number of time steps for the given resolution. Depending on type of the beat $B_x$, the corresponding value $p_x$  of the feature vector $\textbf{p}$ is defined as follows: 
\vspace{-0.1cm}
\begin{description}[noitemsep]
\item[$B_x$ is a significant beat:] $p_x = \sum_i b_i/t$. This dimension represents how frequently that pulse is a high persistent maximum in that resolution. 
\item[$B_x$ is a maxima beat:] $p_x = \sum_i b_i/t$. This dimension represents how frequently that pulse is a maximum in that resolution.
\item[$B_x$ is a function beat:] $p_x = max_i\{b_i\}$. This dimension represents how high the scalar function reaches at that location.
Intuitively, it captures the maximum magnitude of the ``interestingness" a location reaches along the given temporal resolution.
\end{description}
\vspace{-0.1cm}
For a given pulse $P$, its rank is computed as $\|\textbf{p}\|$, the $L_2$ norm of its corresponding feature vector. A high value of rank intuitively implies a high level of activity at the corresponding location.

\begin{figure}
    \centering
    \includegraphics[width=\linewidth]{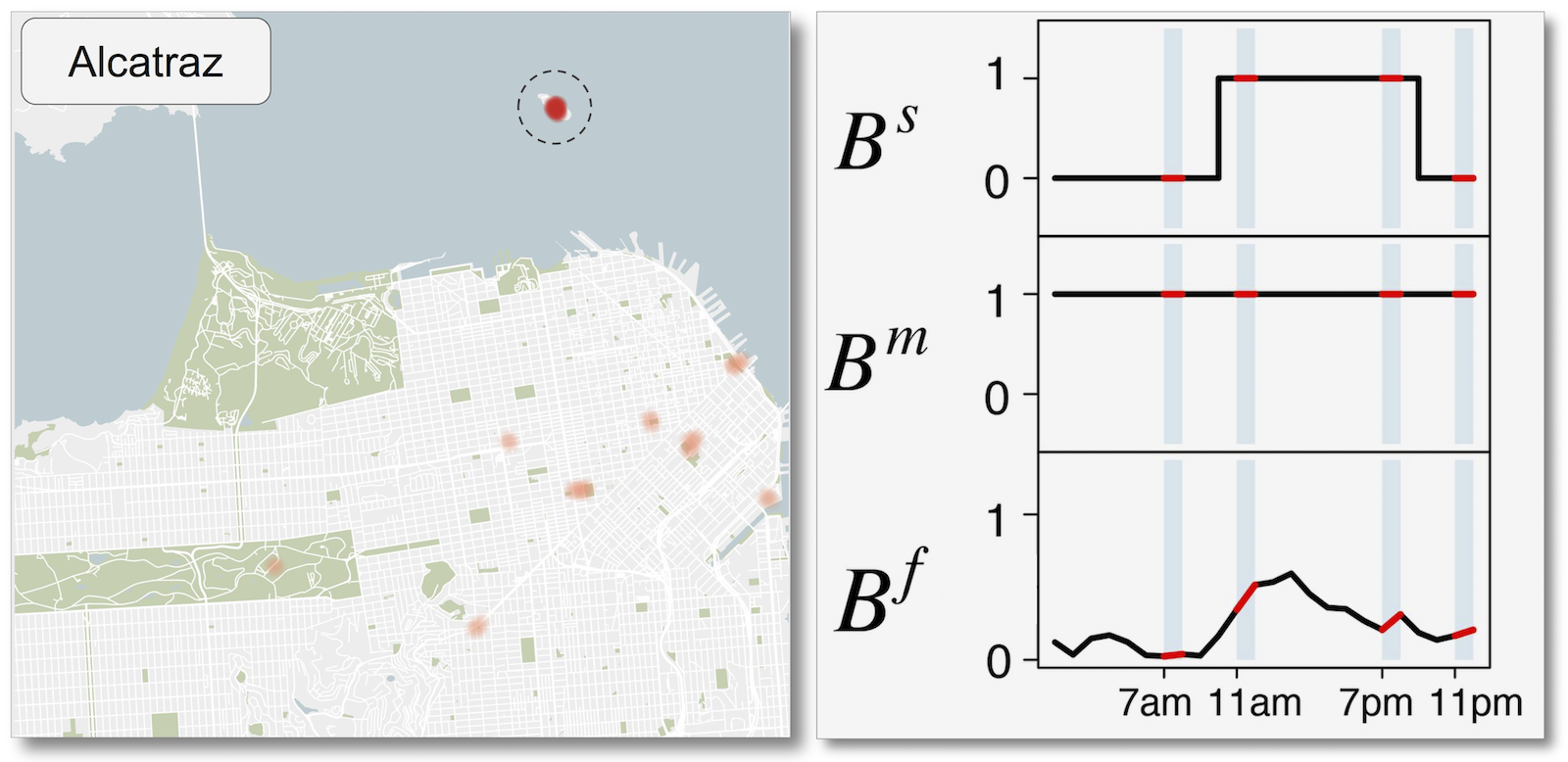}
    \vspace{-0.22in}
    \caption{The different beats of the pulse at Alcatraz island in San Francisco for the hourly resolution.}
    \label{fig:alcatraz}
\end{figure}

\myparagraph{Pulse Similarity.}
A $d$-dimensional similarity vector, $\textbf{s} = (s_1,s_2,\ldots,s_d)^T$, is constructed between a given pair of pulses $P_1$ and $P_2$ as follows. As with the feature vector, each dimension $s_x$ of $\textbf{s}$ corresponds to a beat with respect to a single temporal resolution. The value of the dimension $s_x$ is computed as the Euclidean distance between the two time series $\|B^1_x - B^2_x\|$ corresponding to the $x^{th}$ beat of the two pulses $P_1$ and $P_2$ respectively. 
Each dimension basically represents the similarity between the given pair of beats.
The similarity measure is then defined as the $L_2$ norm $\|\textbf{s}\|$ of the similarity vector. A low value of the measure indicates a high similarity and vice versa. 
For example, consider the pulse corresponding to Alcatraz Island in San Francisco~(SF) shown in \reffig{fig:alcatraz}. When compared to the three locations in \reffig{fig:pulse-beats}, it is most similar (lowest similarity measure) to the Met, followed by Central Park, and Williamsburg. This is because this pulse has more in common with the Met in terms of the different beats. Note that, at the hourly resolution, they have identical maxima beats, and almost identical significant beats. Furthermore, the function ranges, as represented by the function beats, are also closer to each other compared to the other locations.

In case the beats from certain resolutions are missing from one of the two pulses, then the similarity vector is constructed using the common set of resolutions present in both pulses. For instance, such a scenario can occur when the time period of a data set used to create the scalar functions is small, and is not enough to cover all months of a year. In this case, the only resolutions possible are All, Day, and Hour. 

This measure allows us to compare pulses corresponding to different data sets, thus different activities over a city. Furthermore, since the comparison of two pulses is independent of the spatial location, we can also compare pulses across cities. 
As described in Section~\ref{sec:interface}, we use the similarity measure to support querying of similar pulses across different scenarios.

%% file: impl.tex
\section{Implementation}
\label{sec:impl}

In this section, we briefly discuss the various choices made during the implementation of the urban pulse framework.

\begin{table}
    \centering \small
    \begin{tabular}{|c|c|} \hline
    \textbf{Scenario}                       &  \textbf{Resolutions} \\\hline \hline
    Parts of Week (Weekday, Weekend) &  All, Month, Hour \\
    Seasons (Spring, Summer, Fall, Winter)  &  All, Day, Hour \\
    Parts of Day (Morning, Afternoon, Evening, Night) & All, Month, Day \\ \hline
    \end{tabular}
    \vspace{-0.1in}
    \caption{The different scenarios that can be used through our framework to explore and compare pulses within and across different cities.}
    \label{tab:scenarios}
    \vspace{-0.2in}
\end{table}

\begin{figure*}
\centering
\includegraphics[width=\linewidth]{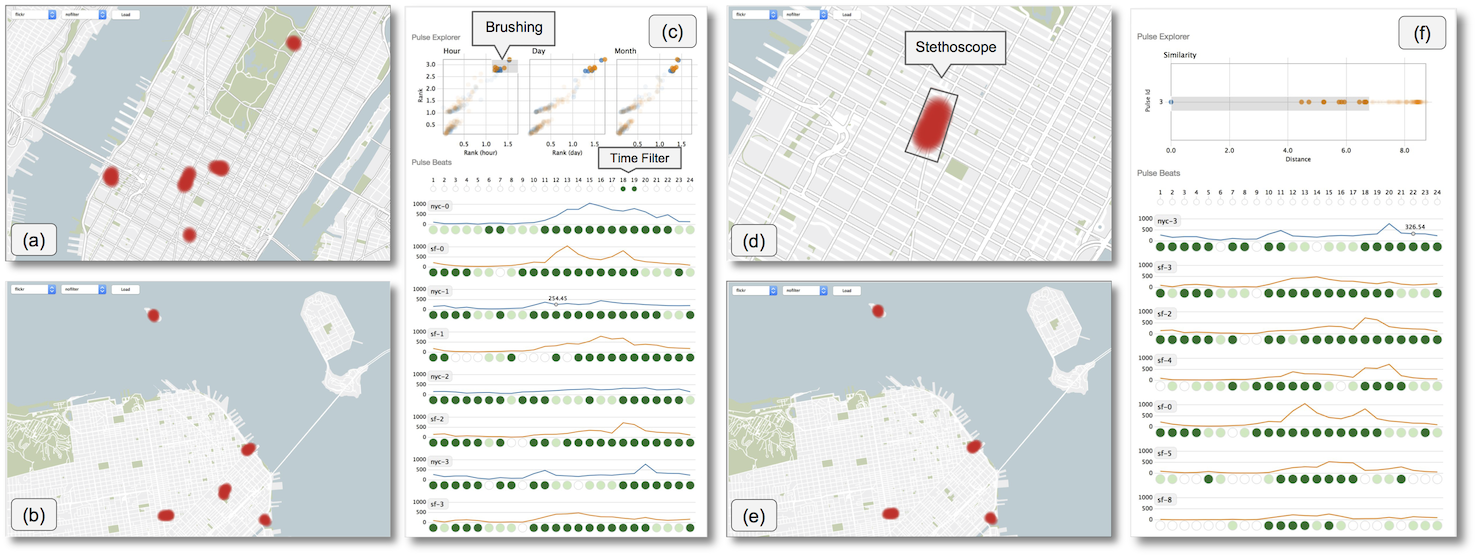}
\vspace{-0.3in}
\caption{Visual Exploration Interface of the Urban Pulse framework. 
\textbf{(a)}, \textbf{(b)}~show two map instances highlighting the highest ranked pulses at the \emph{hour} resolution.
These pulses are selected by brushing the scatter plot in the pulse monitor, shown in \textbf{(c)}. The pulse monitor also visualizes the beats of currently selected pulses. The different colors of the points in the scatter plot, as well as the beats viewer, correspond to the different map widgets (blue for NYC and orange for SF). The shown beats can be filtered using the  time filter. 
The Stethoscope tool is used to select the pulse corresponding to Times Square in \textbf{(d)}. It is then used to query for similar pulses in San Francisco. The scatter plot in the pulse monitor is updated to show the result of the query, shown in \textbf{(f)}. Selecting the closest pulses using the scatter plot highlights these pulses on the map widget, see \textbf{(e)}.
}
\label{fig:gui}
\vspace{-0.2in}
\end{figure*}

\myparagraph{Exploring scenarios.}
The dynamics in a city can vary based on different conditions such as weather, time of day, or day of the week, etc. For example, the activities during summer could be significantly different from those during winter. Similarly, weekend dynamics could differ from weekday dynamics. Experts are interested in not just isolating the activities during different conditions, but also in the comparison between the conditions. 
To accomplish this, we create multiple scalar functions corresponding to each of the possible conditions. Table~\ref{tab:scenarios} lists the different conditions supported in our implementation. Note that each set of conditions is typically a partitioning on one of the temporal resolutions. Hence, there will not be scalar functions corresponding to that resolution. The rank and similarity measures in this case are computed by ignoring this resolution. Table~\ref{tab:scenarios} also lists the temporal resolutions considered for the given set of conditions.

\myparagraph{Mesh representation of a city.}
Given a city, the mesh corresponding to it represents the bounding box of that city. Vertices are sampled uniformly in the form of a grid such that the distance between two vertices (horizontally and vertically) is 50~meters. Our current implementation uses a simple triangulation, which creates two triangles for each grid cell. The meshes corresponding to the two cities used in our use cases -- New York City and San Francisco, had round 142,000 and 136,000 vertices respectively.

\myparagraph{Computing scalar functions.}
All the data sets used in this paper are composed of a set of data points over space and time. While all case studies in the next section use the density functions, the computation of any scalar function would first require identifying points corresponding to each vertex of the input mesh. This boils down to querying for all points within a given neighborhood radius of the vertex, and grouping them based on time. This operation over all vertices is equivalent to a spatio-temporal join between the circular regions defined by the radius and the data points. 
In our current implementation, we use the neighborhood radius $r = 5\times\epsilon$. We use a value of $\epsilon = 100$~m, which is the typical size of a city block.
%
This value ensures that locations with high intensity of activity more than a few blocks away are considered as distinct maxima. While a larger neighborhood radius can be used, it has a smoothing affect on the scalar function resulting in the possible loss of such locations.

Even though the input data can be large, this operation can be performed efficiently using spatial indexes. In fact, we can accomplish this with a single pass over the entire data. In our implementation, we make use of a grid index~\cite{gis-book} for this purpose, as follows. Given the set of vertices of the mesh corresponding to the domain representation of a city, we first build the grid index on the locations of these vertices. A grid index divides the bounding box encapsulating the region of interest into a set of cells, and each vertex will belong to one of these cells. Given the structured nature of the index, the cell corresponding to a given location can be identified in $O(1)$ time. 

The next step iterates over all data points from the data set. For each data point, we first locate the cells within the given radius of its location. Next, the data point is added to the vertices that are part of these cells in the index. 
It is possible that a cell on the border of the circular region defined by $r$ can have vertices that are at a distance greater than $r$. Thus, we check the distance to ensure that the data point is indeed within the required distance of the vertex during this process.
In case of a density function, the weight corresponding to each data point is added to the appropriate time step for that vertex.

We pre-process the data to compute the scalar functions and the pulses, which are then explored using the visual interface. On a laptop having a 2.3 GHz Intel Core i7 processor and 8 GB RAM, it took 18 minutes to create all scalar functions (around 2 seconds per function) for the Flickr data set corresponding to NYC. Note that this includes scalar functions over all temporal resolutions. The identification of the set of prominent pulses took 2 minutes.

%% file: interface.tex
\section{Visual Exploration Interface}
\label{sec:interface}

The purpose of the visual exploration interface was twofold. First, it should allow users to explore pulses with respect to one or more cities. Second, it should support the comparison of pulses not only across different scenarios within a city, but also between multiple cities.
To accomplish this, we design a visual exploration interface that is composed of two components --- 1.~\emph{Map View} that provides spatial context for the pulses; and 2.~\emph{Pulse Monitor} that is used to explore and analyze the different properties of the pulses. These two components consists of a collection of linked visualization widgets together with a set of interaction strategies. \reffig{fig:gui} and the accompanying video demonstrates the different functionality of the Urban Pulse interface.

\subsection{Visualization Widgets}
\label{sec:vis}
We now briefly describe the different visualization widgets corresponding to the two components of the pulse interface.

\myparagraph{Map widget.}
The map view component consists of one or more map widgets and provides the spatial context with respect to the different pulses. The number of map widgets is defined by the user depending on the task. Typically, there are two map widgets to facilitate a two way comparison. 
The set of pulses are rendered as a collections of regions (see \reffig{fig:gui}). The region for each pulse is defined by the corresponding maximum location(s) and the influence radius as illustrated in \reffig{fig:pulse-merge}.
Users can also optionally visualize the scalar function under consideration as a heat map. 

\hidecomment{
This widget contains a vector map over which different visualizations are overlaid. In our current implementation, we support two visualizations on top the map, which the user can selectively enable. 
\vspace{-0.15cm}
\begin{enumerate}[leftmargin=*,noitemsep]
    \item The scalar function that is currently being analyzed is visualized as a heat map with transparency over the city. Recall that the function values are normalized between 0 and 1. Points on the map having very low (close to zero function value) are rendered transparent, while those with a higher function value are more opaque (see \reffig{fig:scalar-fn}). The transparency helps avoid obscuring the map context.
    \item The set of pulses are rendered as a collections of regions (see \reffig{fig:gui}(a), \ref{fig:gui}(b), \ref{fig:gui}(d) and \ref{fig:gui}(e)). The region for each pulse is defined by the corresponding maximum location(s) and the influence radius as illustrated in \reffig{fig:pulse-merge}.
\end{enumerate}
\vspace{-0.15cm}
}

\myparagraph{Linked scatter plots.}
The default mode of the Pulse Monitor consists of three linked scatter plots, one for each of the \textit{hour}, \textit{day}, and \textit{month} resolutions (\reffig{fig:gui}(c)). Each point in the scatter plot corresponds to a pulse in the city. The $y$-axis of these plots represent the rank of the pulses. The $x$-axis represents the rank when restricted to the corresponding resolution. That is, the rank is computed using only the beats corresponding to the given resolution. The scatter plots are linked in the sense that selection of a point (pulse) in one plot also selects the corresponding point in the other plots as well. 
%
The points are colored based on the associated map widget, as shown in \reffig{fig:gui}(c).

\myparagraph{Pulse beat viewer.}
The pulse beat viewer is used to visualize the three beats --- significant beats, maxima beats, and function beats, corresponding to a given pulse and resolution. 
Function Beats are visualized as a line plot where the $x$-axis represents the time step and the $y$-axis represents the function value. The user can choose to visualize either the actual function value or the normalized function value. 
The significant beats and maxima beats are together encoded as a linear collection of colored circles below the line plot. Here, each circle corresponds to a single time step and shares the $x$-axis with the line plot. The circle is light green if it is a maximum, dark green if the maximum is high-persistent, and white if it is not a maximum. 
%
%
This stacked visual representation was inspired by the design of genome browsers~\cite{IGV11}, which also stacks a plot corresponding to the genome binding data with other related information such as reference tracks.

\begin{figure}[t]
    \centering
    \includegraphics[width=\linewidth]{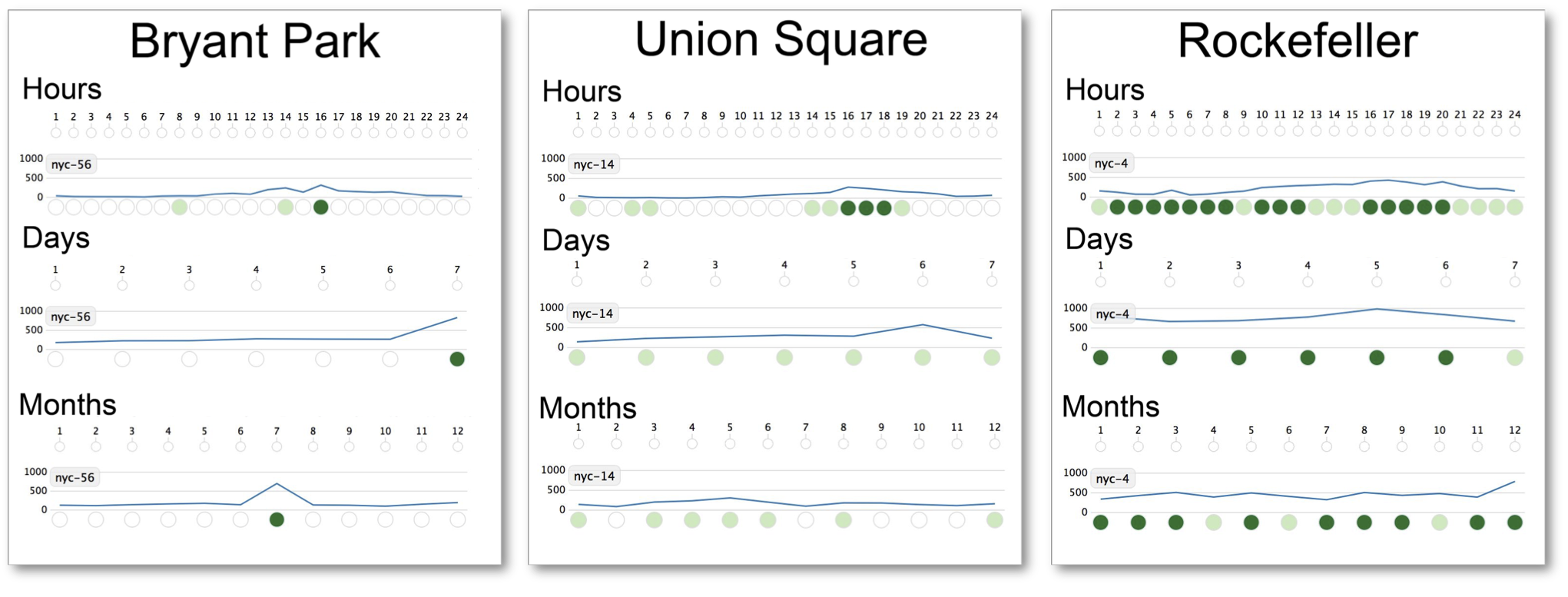}
    \vspace{-0.2in}
    \caption{Hourly, daily and monthly beats for three separate regions: Bryant Park, Union Square and Rockefeller Center. The different roles each space play in the urban ecosystem is clear by looking at the beats at different time resolutions.}
    \label{fig:usecase11}
    \vspace{-0.2in}
\end{figure}

\subsection{Interaction Strategies}
\label{sec:interaction}

We now briefly list the available interaction strategies that can be used to effectively explore and compare different pulses.



\myparagraph{Pulse filtering and exploration.}
Users can select pulses of interest by brushing one of the scatter plots and visualizing their locations on the map. 
%
This operation also results in the corresponding beats being visualized in the pulse beat viewer. The pulses in the beat viewer are sorted based on the rank of the pulses, and alternate between the different data sets being compared. For example, consider the case when the user is comparing the Flickr activity in NYC with SF. 
Then, as shown in \reffig{fig:gui}(c), the beats viewer first displays the top ranked pulse of NYC and SF, followed by the second ranked pulses, and so on. 
In this example, pulses having a high overall rank and high rank in the hourly resolution are chosen using the scatter plot.

Hovering over (or selecting) individual beats also highlights the corresponding pulses on the scatter plots and map widgets.
The beats are rendered with respect to a selected resolution which the users can change. This selection also decides the scalar function that is visualized on the map. 
Users can also choose to use either the normalized or actual values for visualizing  the function beat plot. Furthermore, the beats can also be filtered based on its property at selected time steps. For example, when visualizing the pulses in an hourly resolution, users can select pulse locations that are maximum (or significant) from 11~am to 1~pm. 
This operation is illustrated in \reffig{fig:gui}(c). The color of the filter denotes the filter condition. For example, dark green represents that only pulses, whose locations are high-persistent maxima in the filtered time steps, are considered.

\myparagraph{Spatial selection and querying.}
The interface also supports a \textit{stethoscope tool} (\reffig{fig:gui}(d)), that can be used to select polygonal regions of interest on the map. This restricts the analysis in the pulse monitor to pulses within the given selected region. 

Once a region is selected, it can also be used to query for pulses from other map widgets that are closest to the set of pulses within the selected region. Each pulse from the other map widgets is assigned to the closest pulse from within the selected region. 
%
The search results are listed in the pulse monitor, one below the other, where each line corresponds to one of the source pulses in decreasing order of rank. The similar pulses are ordered horizontally based on the distance (similarity) to the source pulse.
Again, users can brush and select pulses of interest to further analyze the results.

\hidecomment{The results of the query is visualized in a single scatter plot (\reffig{fig:gui}(f)). The $y$-axis of the scatter plot corresponds to the pulses that was queried. The $x$-axis corresponds to the pulse similarity measure. Users can then select pulses of interest and view their corresponding beats and locations. This operation is illustrated in \reffig{fig:gui}(f), and the locations of resulting pulses in SF are shown in \reffig{fig:gui}(e). 
We decided to use a scatter plot for exploring the query results in order to have a consistent interface for all operations.}

%% file: application.tex
\section{Case Studies}
\label{sec:app}

A prototype implementation of the urban pulse framework was used by multiple users specializing in different domains. In this section, we briefly discuss the different case studies carried out by them.

\subsection{Urban Planning}
\label{sec:architects}

In this section we, the architects and urban designers, present two use cases using the urban pulse framework to analyze public spaces for use as precedents in the design process. By combining a more nuanced understanding of the spaces with the discovery of unexpected relevant precedents, we can define new classifications of public spaces.

\begin{figure}[t]
    \centering
    \includegraphics[width=\linewidth]{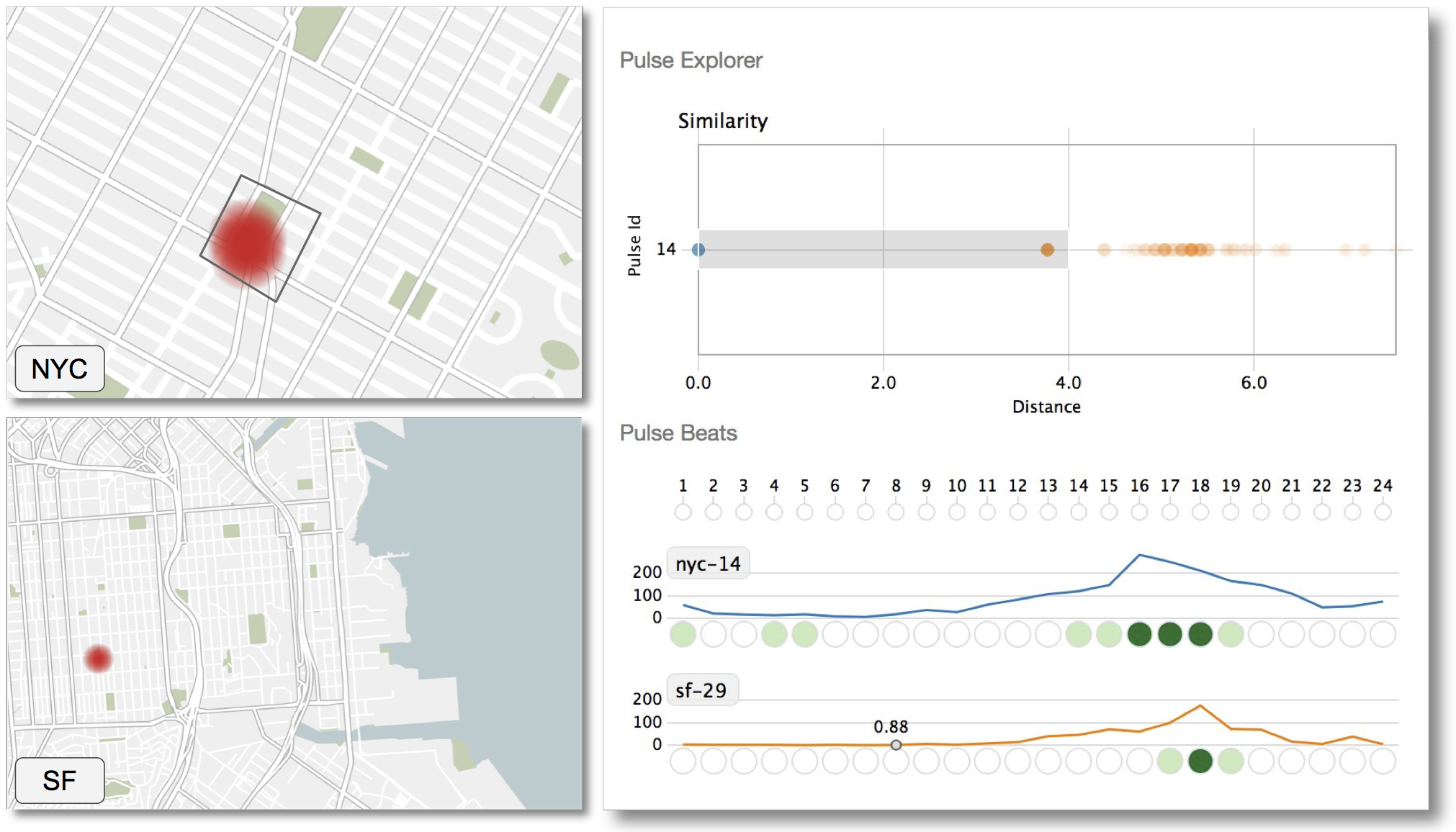}
    \vspace{-0.2in}
    \caption{Using the pulse similarity, it's possible to find analogs to famous NYC landmarks. Bryant Park, for instance, is closely related to Mission Dolores Park. This data driven space characterization can significantly change how public spaces are designed.}
    \label{fig:usecase12}
    \vspace{-0.2in}
\end{figure}

\begin{figure*}[t]
    \centering
    \includegraphics[width=\linewidth]{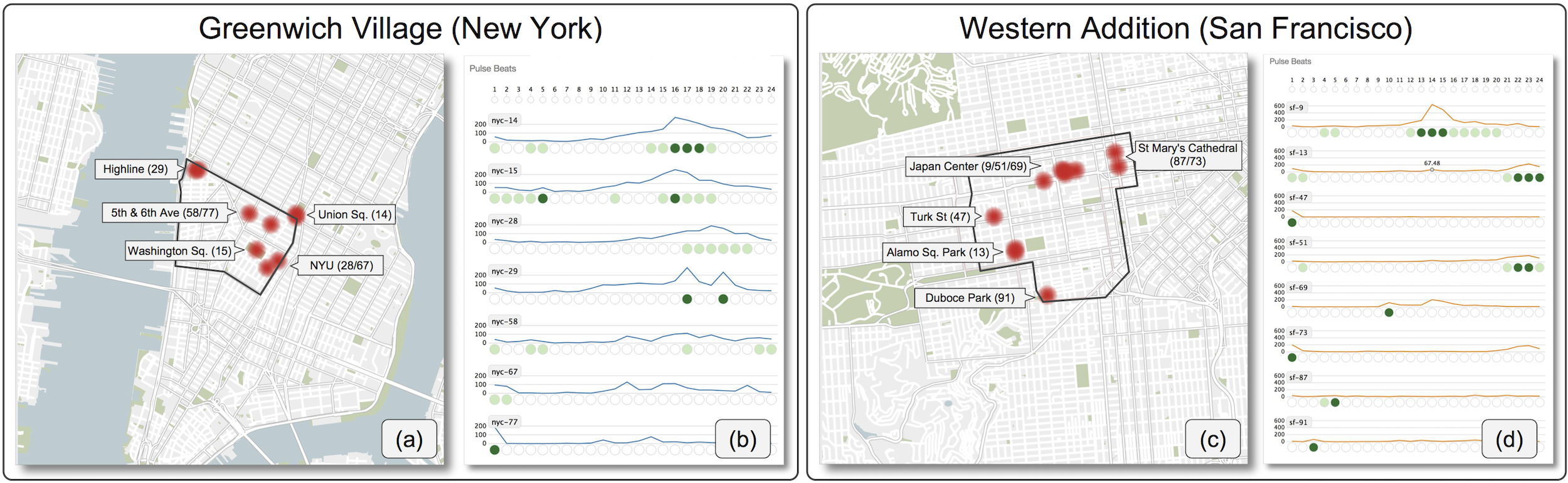}
    \vspace{-0.25in}
    \caption{Understanding the characteristics of a neighborhood. 
    Locations and beats of the prominent pulses corresponding to two famous neighborhoods in NYC and SF indicate a stark difference between their behavior. While Greenwich Village satisfies general expectations, the same is not true with respect to The Western Addition.
    }
    \label{fig:usecase2}
    \vspace{-0.25in}
\end{figure*}

\myparagraph{Characterizing Design Precedents.}
Commonly used spatial precedents in design rely on subjective properties such as land use, program, and density to classify similar spaces, however the interactions of these properties are highly complex and change from site to site. The proposed framework presents a data driven approach to urban space categorization, allowing designers and planners to search for site analogs based on recorded use patterns, revealing groupings of similar sites that defy traditional classification. 

In the first case study we examine three seminal public spaces in New York City that are commonly used as precedents for the design of new urban spaces: Bryant Park, Union Square, and Rockefeller Center. These spaces are normally categorized together because of their similar land uses, contextual densities, size, management, and are all considered successful.
Using urban pulse we find that these illustrious urban spaces, despite all being high intensity pulse locations, have \textit{very little similarity} with regards to their use patterns as defined by Flickr activity between January 2012 and July 2014.
\reffig{fig:usecase11} shows the different beats of the pulse at these locations over all temporal resolutions.
Rockefeller Center's pulse is consistent (is a significant maximum at most time steps, and at least a maximum at others) across all three resolutions, the closest NYC analogue being Times Square. 
There is significant activity in Bryant Park primarily in the evening (4pm). In the daily and monthly resolutions, Bryant Park is significant on Sundays and in July. 
The pulse of Union Square, on the other hand, shows that it is more popular during the second half of the day, with it becoming significant towards work closing time (4pm-6pm). The pulse is consistently a maximum on all days even though it is not a significant maximum, although one notices more activity on Saturdays by looking at the function beat. Similarly, we notice that the pulse is consistent during the warmer months of the year.
Even when using only a single data set we already see nuanced differences in these urban spaces that a subjective traditional categorization would miss. The different roles these spaces play in the urban ecosystem is a little clearer, Rockefeller Center is a high intensity tourist destination, and while Union Square caters to a steady flow of local users, it is still relatively popular with tourists compared to Bryant Park. Bryant Park represents a mix, with high use throughout the year from surrounding offices (and therefore less Flickr activity), but intense programmed spikes on weekends or in summer due to the happening events. 
Using the pulse similarity to find these space's Flickr analogues, we find Union Square to be very similar to Washington Square Park, while Bryant Park has no true equivalent.

\hidecomment{
For example, Rockefeller Center's pulse is consistent across hourly and daily resolutions, the closest NYC analogue being Times Square. Bryant Park's Pulse is a highly concentrated beat at 4:00pm on Sundays in July. Union Square's pulse is a long daily beat, centered at around 4:30pm in the Spring months with relative increases in intensity on Saturdays. Even when using only a single data set we already see nuanced differences in these urban spaces that a subjective traditional categorization would miss. The different roles these spaces play in the urban ecosystem is a little clearer, Rockefeller Center is a high intensity tourist destination, Union Square caters to a steady flow of local users. Bryant Park represents a mix, with high use throughout the year from surrounding offices, but intense programmed spikes on weekends corresponding to summer events. Using the pulse similarity to find these space's Flickr analogs, we find Union Square to be very similar to Washington Square Park, while Bryant Park has no true equivalent.
}

Similarly, spaces can be identified in other cities as well such as San Francisco. Taking the pulse of Bryant Park reveals a striking similarity across all scales to Mission Dolores Park~(\reffig{fig:usecase12}). The pulse at Union Square finds a hourly corollary to Mckinley Square in Portrero Hill. And interestingly Rockefeller Center finds a daily scale analogue with Alcatraz. Although no one site may have an exact twin in other cities, we can learn from the way similar sites differ. 
These comparisons allow architects to learn from the success or failure of spaces that have similar occupation but would otherwise not have been considered. For example, if the design or operation of Alcatraz needed changing, we could look to Rockefeller Center for reference.

These new analogues defy currently understood use relationships and indicate the emergence of a data driven method of characterizing places. This results in a new category of spaces characterized not by their physical characteristics but by their actual occupation by people. These new categories can significantly change how we design new public spaces.

\myparagraph{Understanding Neighborhoods.}
For architects, planners, and urban designers, neighborhood activity patterns and points of interest are the result of intensive ethnographic surveys that take years to conduct and given the speed at which neighborhoods change, can be out of date quickly. The features provided by the urban pulse framework helps examine the diversity, distribution, and intensity of human activity within a given neighborhood, thus offering insight to the functioning of the entire neighborhood. For this study we examine two neighborhoods known for their distinct character: Greenwich Village in New York, and The Western Addition in San Francisco.

Greenwich Village is often seen as quintessentially New York in urban morphology and character. Home to the central campus of New York University (NYU) and Parsons The New School, it is a locus of student activity and provides many destinations for tourists and locals alike. Initial analysis of Greenwich Village exposes seven prominent pulses based on Flickr activity, see \reffig{fig:usecase2}(a). The locations of these pulses are centered around noted landmarks -- NYU, Union Square, The Highline, The Washington Square Arch, and points around the NYU campus on 5th and 6th Avenue. 
The three highest ranked pulses indicate high activity during the second half of the day~(\reffig{fig:usecase2}(b)). 
%
%
Filtering based on time of day (Table~\ref{tab:scenarios}) showed some variation in the location of pulses within the neighborhood, but in general the patterns remained consistent and stable.

The Western Addition is a diverse neighborhood with several internal boundaries created throughout its history and represents much of San Francisco's middle to high income residents. It is composed of major public spaces like Alamo Square Park, and landmarks such as surviving historic Victorian row houses, St. Mary's Cathedral, and Japan Center. Initial analysis using Flickr activity indicated eight prominent pulse locations in this neighborhood (\reffig{fig:usecase2}(c)). The four highest ranked pulses correspond to a location in Japantown Center, a location to the West of Alamo Square Park on Divisadero, a location on Turk Street, and Duboce Park. In contrast to Greenwich Village, only Japan Center, which is active  from 12:00pm to 8:00pm, and Alamo Square Park (at nights) have any bustle here. As seen in \reffig{fig:usecase2}(d), there was a distinct lack of strong beats at major points of interest, which is counter-intuitive to conventional wisdom.
However, higher activity was observed at landmarks such as Alamo Square Park and St. Mary's Cathedral when exploring the pulses during the weekend. 

The preliminary analysis performed above, by itself, helps draw several useful conclusions about neighborhoods and their specific regional assets. The Flickr activity pattern for Greenwich Village indicates confirmed activity in the afternoon hours and can influence the decision to program Washington Square Park accordingly.
Conclusions with respect to The Western Addition proved more complicated. %
Despite status as landmarks, these locations had very weak pulses, indicating a needed change in planning strategy. 
Both the above cases revealed locations of interest that would otherwise not have been possible from a traditional strategy that only considers locations based on physical characteristics and land use patterns. This is primarily because the latter would not include \textit{unclassified regional assets} -- locations that are not known landmarks but still attract activity.
Of particular importance to urban planners and designers is the emergent quality of these locations that aren't normally found outside of intense ethnographic study, and the additional information urban pulse provides about their temporal appearance. 


\subsection{Behavioral Patterns of Cultural Communities}
\label{sec:twitter}

In our increasingly globalized world it is becoming more and more common for individuals from widely different backgrounds and cultures to find themselves living side by side within the same urban space. The way in which this cohabitation takes place, how common spaces are shared and how the different communities interact both directly and indirectly is far from understood. 

We now take the first steps at a qualitative approach to this problem. 
We use the set of geo-located Tweets produced between May $2010$ and December $2015$ for this study. We first classify the set of tweets based on language of the tweet~\cite{mocanu13-1} and the pulses are computed for each language. That is, we use language of the tweet as a proxy to represent the activities of the corresponding community. 
While the raw number of tweets does not necessarily correspond to the number of individuals in the underlying population, we expect that the way in which these users behave as seen through the Twitter lens is representative of the behaviours of that community. 

For simplicity, we start by focusing on the comparison between English and Spanish tweets as they are by far the largest communities present in NYC and hence better represented in our data set.
While the overall temporal and geographical distribution of Twitter activity in each of these languages is highly inhomogeneous and complex, the Urban Pulse approach is still able to identify the most significant locations, across various timescales, even when one of the signals is considerably larger than the other. This is a major advantage over current approaches! As a first step in exploring the differences between the two groups, let us take a close look at the Evening and Night periods. 

\begin{figure}[t]
    \centering
    \includegraphics[width=\linewidth]{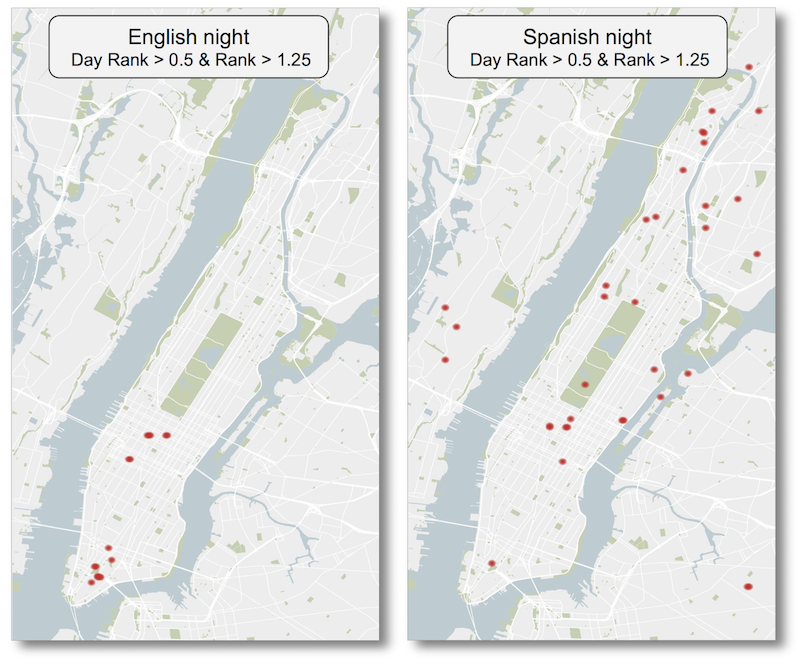}
    \vspace{-0.25in}
    \caption{Nighttime pulses show a clear divide on the locations of English-speaking and Hispanic activity in NYC.
    }
    \label{fig:usecase4a}
    \vspace{-0.25in}
\end{figure}

During these periods, the most significant English pulses are exclusively located in the area below Central Park, while the majority of Spanish pulses are located in the Harlem and Bronx areas (see \reffig{fig:usecase4a}). A pattern that is valid everyday of the week and that matches well with our perceptions of the demographics of these areas, serving as further validation of this technique. The Harlem and Bronx areas have historically had a large Hispanic population and one might expect that users are more likely to tweet from their home neighborhoods during the night. Similar patterns are detectable during the rest of the day, although much less pronounced and with activity more evenly distributed. A similar contrast can be observed between Portuguese and French speakers, with French speakers, mostly originating from former French colonies concentrating in the northern areas while Portuguese speakers are more active south of Central Park. 

New York is also famous for its large Italian-American population that has been part of the city's history for over a century. This level of integration makes it less likely for them to tweet in their own language and might explain why, despite these historical connections, Italian is only the $9^\textrm{th}$ most common language in our data set. A deeper look at the Italian data layer makes clear why this is the case. Significant Italian pulses are overwhelmingly concentrated in world famous touristic attractions such as the World Trade Center memorial, the Empire State Building, Times Square, Central Park or the Flat Iron building, indicating that they likely originate from tourists, who are known to behave differently from local residents~\cite{bassolas16-1,humandiffusion}. This is further reinforced when we compare Weekends and Weekdays, which show no significant difference. During Summer, in addition to the touristic locations mentioned above we also see significant pulses along the water line that are absent during the Winter period. Similar patterns can also be observed in other languages without an historical connection to the city, such as Indonesian or German.

As a further comparison, we consider the case of the Korean speaking population. This community also has a long history in the city, with a large ethnic neighborhood, Korea-Town, on 32$^{\textrm{nd}}$ St. between 5$^{\textrm{th}}$ Ave. and Broadway. In this case, one of the most significant pulses is precisely in the vicinity of Korea-town. Also, there is no clear difference between Summer and Winter indicating that the majority of the activity in this layer is due to residents instead of tourists.

%% file: conclusion.tex
\section{Discussions and Future Work}
\label{sec:disc}

\myparagraph{Expert Feedback.}
Throughout the research and development of the framework, we kept close contact with the domain experts, tuning the interface and data exploration to satisfy their needs. Once they became familiar with the workings of the software, we requested feedback on the following aspects of the framework -- utility, ease of use, new feature requests, and their plans for future usage.
%


All users agreed that the system was \textit{incredibly useful}, allowing them to quickly detect non-trivial behavioral patterns over the city.
Architects, who have had prior experience with similar systems, found the interface \textit{intuitive} to use. In particular, they liked the linked map and the scatter plots, which according to them ``made it very easy to explore spaces and understand activity levels".
On the other hand, the human behavioral expert had initial difficulties with the interface, but mentioned that the \textit{experience quickly improved} as the interface tuning progressed. This user also appreciated the fact that the ``tool enables the possibility to quickly identify differences between the way in which two areas of the city are used at different times or according to different data layers, allowing for a quick construction of a mental picture of how each area fits within the human and cultural landscape of the urban fabric."
Both sets of users had suggestions on usability enhancements as well as new feature requests. Common among them was the need to support an easy mechanism for computing and loading alternate scalar functions. We are currently in the process of incorporating these requests into the framework.
The architects plan to use this framework in their ongoing and future projects involving the design of public spaces, while the human behavioral expert intends to expand his study by involving multiple other geo-located temporal data sets.

We would like to note that the use cases were performed independently by the domain experts without any supervision by the visualization experts. 
We would also like to stress that the main focus of this work was in the application of topology-based techniques to solve the problem of understanding and comparing cities using available urban data sets. 
While usability plays an important role in this, we plan to address this through a detailed study in the future.

\myparagraph{Using other urban data sets.}
Our current implementation assumes that the data set consists of a set of points each consisting of spatial and temporal attributes. We would like to note that this is a indeed the most common format for spatio-temporal urban data sets. For example, NYC alone has released over 1300 data sets of which several data sets have spatio-temporal attributes in the above format~\cite{barbosa2014structured}.
Publicly available geo-tagged social media data sets also follow this format.
Thus, any of these data sets can be used with our current implementation as is. 
Other spatio-temporal data sets from NYC open data provide data corresponding to pre-defined partitions of the city such as neighborhoods and zip-codes. In such cases, we use the existing segmentation itself as the mesh.
While we currently do not use other attributes of the data (such as tweets, or Flickr images), it will be interesting to explore ways to transform such data into scalar functions.

\myparagraph{Computing alternate scalar functions.}
In this paper we focused mainly on the density function. In case an alternate function is to be computed on the urban data, then that would essentially require computing the appropriate measure on each data point, and aggregating these measure based on time and influence region of the vertices of the input mesh. It is straight forward to extend the algorithm described in Section~\ref{sec:impl} for this purpose.
For example, consider the twitter data set. Let the user be interested in the average sentiment for each vertex. Then instead of using the density of this data point during the computation, the sentiment value is computed and added to the influencing vertices at the appropriate time step. Additionally, a counter is maintained which keeps track of the number of data points with respect to each mesh vertex and time step. After the input data is processed, then the average sentiment for each vertex is equal to the computed sum of sentiment values divided by the counts.

\myparagraph{Threshold for identifying prominent locations.}
We used a low value for the threshold (0.2) to define a high persistent maximum. This could potentially create a higher number of prominent pulses.
However, most pulses, having low persistence value also have a low rank. 
Since in most cases, users are interested in a higher ranked pulses, these can be easily filtered using the scatter plot as they are typically located near the origin.
Also, being conservative helps in not missing out on an interesting pulse.
Our framework also allows users to define the threshold.

\myparagraph{Future work.}
Currently users typically explore pulses one data set at a time. However, it will be interesting to create combined pulses taking into account multiple data sets. One way to accomplish this would be to expand the feature vector for each location to include beats with respect to different data sets, thus increasing the dimension of the feature vector. This would also enable the use of the current user interface without any modifications.
Alternatively, it will also be interesting to explore multi-variate techniques to characterize combined pulses.
So far, the pulse was defined based on the maxima of a scalar function. In future, we plan to generalize the definition of a pulse to be based on one or more critical point types.

\hidecomment{
\myparagraph{Definition of Pulse.}
Since the functions used in this paper was the density function, it was suitable to use the maxima of the scalar functions to identify and characterize pulses. Depending on the scalar function used, the maximum might not always be the correct choice. However, since topological persistence is defined for all critical point types,  For example, if a minimum is used, then the prominent locations will be identified based on the locations on high-persistent minimum. 


Should there be a discussion on the merging of locations??

}